# Three-dimensional laser printing of macro-scale glass objects at a micro-scale resolution

Peng Wang,[1,2,3] Wei Chu[2,6] *, Wenbo Li,[2,4] Yuanxin Tan,[2,3] Fang Liu,[4] Min Wang,[5,6] Zhe Wang,[2,4] Jia Qi,[2,3] Jintian Lin,[2] Fangbo Zhang,[2,3] Zhanshan Wang,[1] Ya Cheng[2,5,6,7, †]

[1]School of Physics Science and Engineering, Tongji University, Shanghai 200092, China. [2]State Key Laboratory of High Field Laser Physics, Shanghai Institute of Optics and Fine Mechanics, Chinese Academy of Sciences, Shanghai 201800, China. [3]University of Chinese Academy of Sciences, Beijing 100049, China. [4]School of Physical Science and Technology, ShanghaiTech University, Shanghai 200031, China. [5]State Key Laboratory of Precision Spectroscopy, School of Physics and Materials Science, East China Normal University, Shanghai 200062, China. [6]XXL - The Extreme Optoelectromechanics Laboratory, School of Physics and Materials Science, East China Normal University, Shanghai 200241, China. [7]Collaborative Innovation Center of Extreme Optics, Shanxi University, Taiyuan, Shanxi 030006, China.

Correspondence and requests for materials should be addressed to W. C. (email: chuwei0818@qq.com) or to Y. C. (email: ya.cheng@siom.ac.cn).

**Three-dimensional (3D) printing has allowed for production of geometrically complex 3D objects with extreme flexibility, which is currently undergoing rapid expansions in terms of materials, functionalities, as well as areas of application. When attempting to print 3D microstructures in glass, femtosecond laser induced chemical etching (FLICE) has proved itself a powerful approach. Here, we demonstrate fabrication of macro-scale 3D glass objects of large heights up to ~3.8 cm with a well-balanced (i.e., lateral vs longitudinal) spatial resolution of ~20 μm. The remarkable accomplishment is achieved by revealing an unexplored regime in the interaction of ultrafast laser pulses with fused silica which results in aberration-free focusing of the laser pulses deeply inside fused silica.**

1. Introduction

Nowadays ultrafast lasers (i. e., picosecond lasers of pulse durations below tens of picoseconds and femtosecond lasers) have been widely adopted in fabricating three-dimensional (3D) microstructures in various transparent materials[1-6]. In particular, ultrafast lasers have enabled fabrication of geometrically complex 3D microstructures in glass for a variety of applications ranging from microfluidics and microoptics to micromechanics and 3D printing. Generally speaking, ultrafast laser pulses with sub-ps durations are considered more advantageous than the picosecond laser pulses in terms of the highest achievable spatial resolution as well as energy deposition efficiency as the shorter the laser pulse durations, the less significant the thermal diffusion and the stronger the interaction of the laser pulses with the materials owing to the enhanced peak intensities[7]. For instance, when combining the focusing power of high numerical aperture (NA) lenses and the extreme sensitivity in the response of large bandgap glass materials to the laser intensity, the femtosecond laser pulses have enabled to achieve nanoscale resolution in both surface and in-volume structuring of glass[8, 9]. The facts seemingly encourage to choose only high-NA objectives for high resolution 3D structuring of glass owing to a rapid degradation of the axial resolution with the decrease of the NA. With the diffraction of light waves, the axial resolution is inversely proportional to the NA of the focal lens in the linear interaction regime, and the resolution can be further spoiled when the nonlinear self-focusing of the pulses is taken into account[10].

Unfortunately, the high-NA focal systems are inherently of short working distances. Thus, the depth of focal position in the glass allowed by the high-NA lens is typically limited to a few millimeters, which is often desirable to be significantly extended. Previously, a simultaneous spatiotemporal focusing (SSTF) scheme was proposed to tackle this problem, which is able to maintain a reasonable axial resolution even with low-NA focal lenses[11, 12]. Recently, this scheme has enabled 3D micro-printing of fine-featured objects in polymer with heights up to 1.3 cm[13, 14]. Nevertheless, implementation of the SSTF scheme adds extra complexity and cost in femtosecond laser 3D micromachining. Here, we show an amazing and unexpected finding that a nearly spherical modification volume can be produced at arbitrary depths within fused silica with a low-NA focal lens by properly chirping the femtosecond laser pulses into picosecond laser pulses. In other words, one can now achieve aberration-free focusing using long working distance focal lenses regardless of the focal position within fused silica. This unique characteristic provides an opportunity to accomplish laser printing of macro-scale 3D objects in glass at a micro-scale resolution in an easy and flexible manner, which has never been achieved in FLICE despite the great effort spent on it over the past two decades.

## 2. Aberration-free focusing in fused silica with loosely-focused picosecond laser pulses

To examine the fabrication resolutions offered by loosely focusing the picosecond laser pulses into fused silica as a function of focal position along the propagation direction, we first inscribe multiple lines as schematically illustrated in Fig. 1(a). The lines, which are inscribed at the different scan speeds and depths within fused silica, are organized into two grid arrays oriented in X and Y directions, as shown in Fig. 1(b) and (c), respectively. The scan speed and the depth of the focal position can be identified for each inscribed line as indicated in Fig. 1(b) and (c). The average laser power is fixed at 1.65 W in the writing of the lines in both of the two arrays, and the laser pulses are always polarized along Y direction. The results in Fig. 1 are obtained using laser pulses negatively chirped to 10 ps. However, as shown in Fig. S1 in Methods, the same results can also be obtained if one chooses to use positively chirped pulses as long as the pulse duration is properly chosen.

We notice that the cross section of all the inscribed lines shows a similar geometry of nearly circular shape, which are insensitive to the scan speed, depth of focal position as well as the direction of laser writing. The difference is that with an increasing scan speed, the color in the cross section captured under the microscope in a reflective mode becomes lighter, indicating that a weaker modification of fused silica will be generated with the decrease of the irradiation dose at the increasing scan speed. Since in the 3D glass printing with the FLICE, a chemical wet etching must be carried out for selectively removal of the modified fused silica by the laser pulses, it is important to examine the cross section of hollow channels produced after the chemical wet etching in KOH solution, as shown in Fig. 1(d). The lateral and axial resolutions revealed by all the cross-sectional micrographs in Fig. 1(d) are ~20 μm, despite the fact that the lines are inscribed at various depths across a range of 5 cm. A high scan speed of 40 mm/s is chosen for inscribing the lines in Fig. 1(d), indicating that the aberration-free focusing of picosecond pulses is suitable for high throughput manufacturing of 3D objects in glass.

Another unexpected observation in the FLICE with the laser pulses of ~10 ps is that nanograting formation can be avoided as we have discussed in another recent publication[15]. It is well known that under the femtosecond laser irradiation, nanogratings tend to form inside various transparent materials including glass materials such as fused silica as well as several crystals[16, 17]. The mechanism is still under debate whilst this effect has been identified to play significant roles in the applications of microfluidics and photonics[18, 19]. Nevertheless, the nanograting leads to an etching selectivity sensitively depending on the orientation of the polarization of the writing laser beam, which increases the complexity of the beam steering system due to the requirement on the dynamic control of the polarization orientation in the 3D glass printing. Because we can eliminate the formation of nanogratings in fused silica with the picosecond pulses whilst still maintain the highly selective etching as shown in Ref. 15, we are able to perform 3D laser printing without the need of manipulating the polarization of the writing laser beam in real time. This greatly simplifies the beam steering in the printing system, making the whole printing process more robust and easier to put into practice.

### 3. 3D printing of macro-scale objects in glass

The optimum inscription condition determined in Fig. 1 allows us to print macro-scale 3D objects at micro-scale resolutions. Figure 2 shows an Albert Einstein's head sculpture with a height of 1.8 cm. The micrographs shown throughout this work were obtained using an optical microscope (Olympus BX53). The model of the sculpture is presented in Fig. 2(a). Figure 2(b-e) shows the front, right, back and left sides of the sculpture to exhibit the details from the different angles of view. In particularly, the fine features on the face of the great physicist, including the wrinkles, the beard, and the eyelids, are all clearly visible in Fig. 2(f) and (g). It proves that the entire sculpture is printed with a decent fabrication resolution from top to the bottom.

Figure 3 shows another sculpture of a greater height of 3.8 cm, which is a statue of Confucius. Again, the model of the sculpture is shown in Fig. 3(a), and the printed sculpture is presented from the different angles of view in (b-e) in the order of (b) front, (c) left, (d) back, and (e) right sides. The details of the decorative pattern on the cloth, the right side of his face, and the left hand hanging behind his body are shown in the insets on the right-hand side of the images in (b), (c) and (d), respectively. The surface of the whole sculpture appears smooth although it does not reach the level of the mirror-like surface quality typically produced with mechanical polishing. Improvement on the surface quality is possible with elaborated post-annealing or $CO_2$ laser polishing, which requires much effort to optimize and will be investigated in the future.

At last, we demonstrate an air turbine with movable parts directly fabricated within glass without any assembling process. The model of the air turbine is illustrated in Fig. 4(a) and (b). As shown by

Fig. 4(b), the micromachine is composed of an air fan, one driving gear and two driven gears, as well as two cams. The driving gear is fabricated with the turbine fan as an integral component to ensure a robust physical connection between the gear and the turbine fan. The two driven gears can be wound by the driving gear when air flow drives the fan to rotate. Each of the driven gears is connected with a cam. The laser printed air turbine is presented in Fig. 4(c). The micro-machine is functional as we can rotate the cams by injecting an air flow from the inlet. The synchronized motion of the two cams are evidenced by the images in Fig. 4(d) and (e), in which the orientations of the cams are changed as indicated by the white arrows. Thanks to the capability of fabricating large objects at the high resolution, the demonstrated technique offers potential for manufacturing precision instruments, tools and machines in various research and application fields.

4. Conclusion

To conclude, we demonstrate 3D laser printing of glass-based macro-scale objects with heights up to ~4 cm at a resolution of a few tens of micrometers. With the scan speed reaching 40 mm/s and the layer spacing being set at 50 μm as we demonstrate in the current experiments, the fabrication efficiency can be determined to be 0.16 $mm^3$/s. Further improvement on the printing efficiency will be done in the near future by combining a 2D galvo scanner with the 2D motion stage. This design will allow both a high printing speed and a large printing area. The novel 3D glass printing technique is established based on two unconventional characteristics in the interaction of loosely focused picosecond laser pulses with fused silica, namely, the depth-independent aberration-free focusing and the elimination of the self-organized nanograting. The physical mechanisms behind these interesting effects have not yet been clarified. We stress that the interaction of ultrafast laser pulses with transparent media under the loose focusing condition is a largely unexplored area of research, which shall inspire significant interest for further investigations. The high-resolution 3D printing of macro-scale objects in glass is expected to have implications in the fields of photonics, microfluidics, and high-precision mechanics.

**Figures and figure legends**

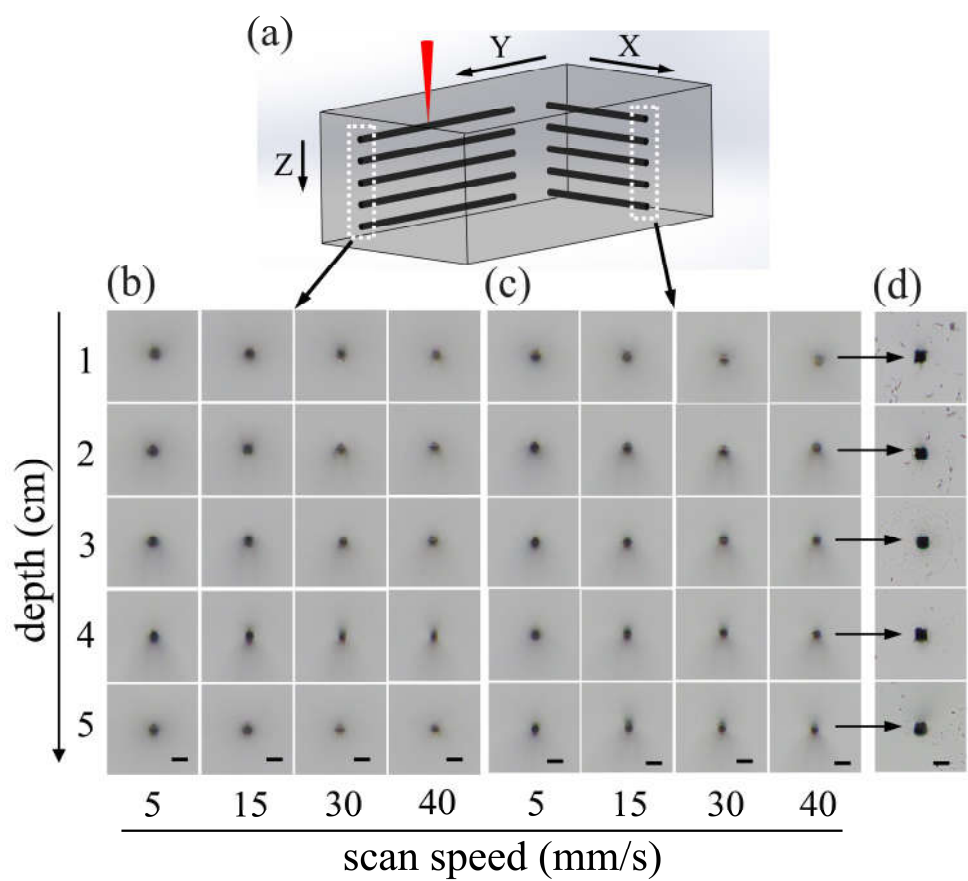

Fig. 1 The fabrication resolution offered by loosely focusing the picosecond laser pulses into fused silica. (a) Schematic illustration of inscribing lines within a cube of fused silica along X and Y direction. Cross-sectional optical micrographs of the lines written along (b) Y and (c) X directions. (d) Cross-sectional micrographs of the hollow channels produced by chemically etching the inscribed sample in the last column of (c). Scale bar: 25 μm.

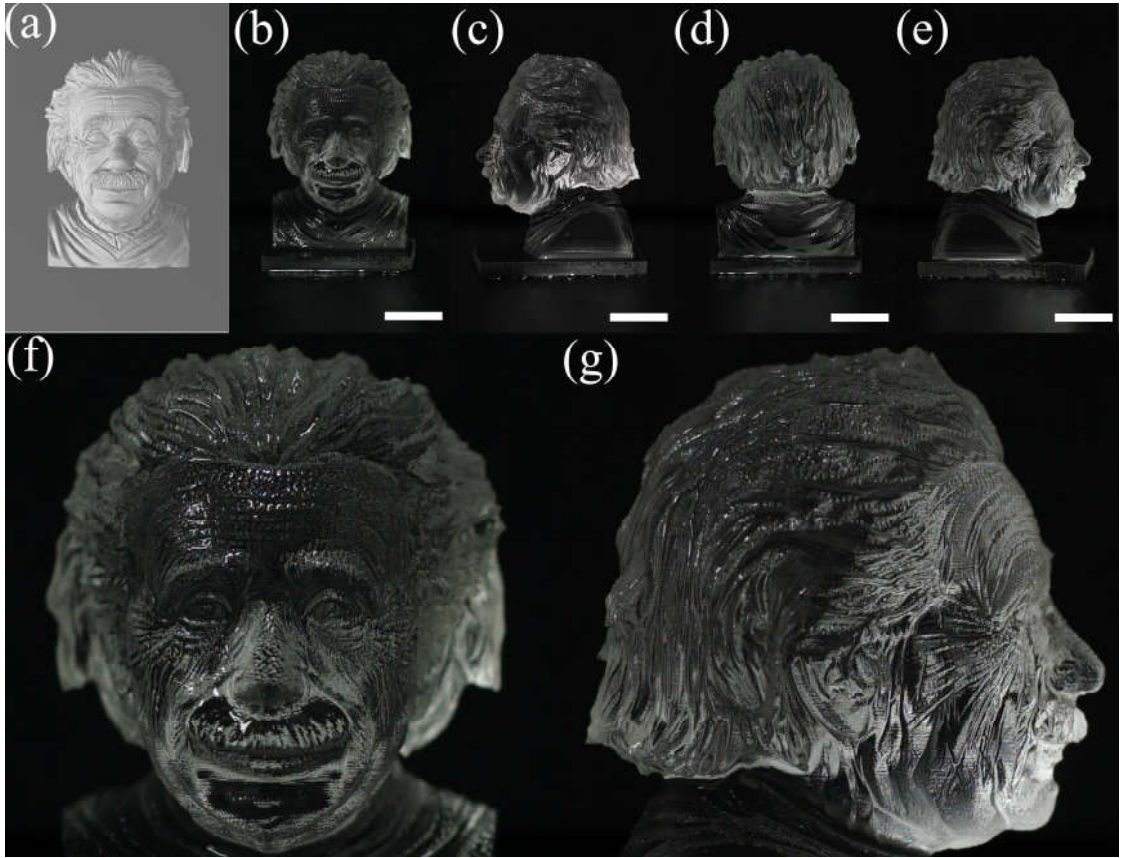

Fig. 2 A laser printed sculpture of Albert Einstein's head in fused silica. (a) The model and the (b) front, (c) right, (d) back, and (e) left sides of the sculpture. (f) and (g) are the zoom-in images of (b) and (e), respectively, to highlight the fine details on the face. Scale bar: 5 mm.

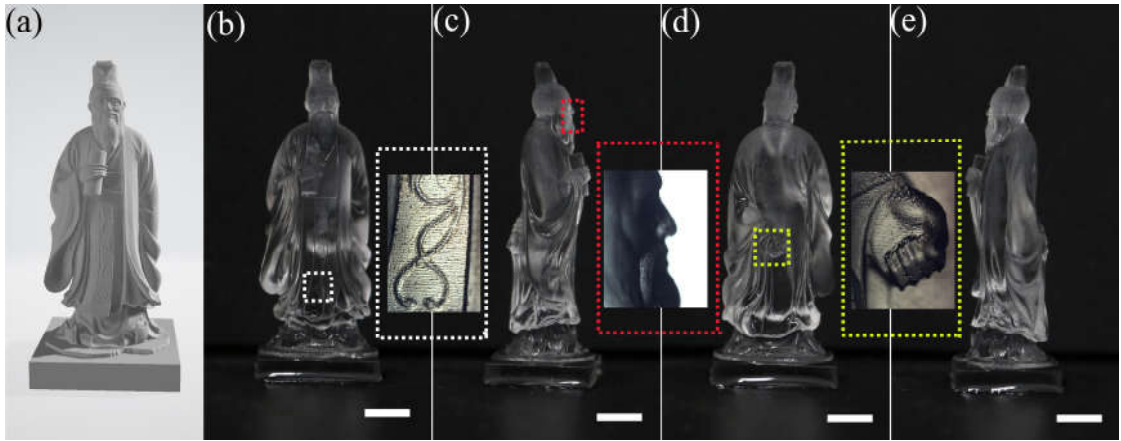

Fig. 3 A laser printed sculpture of Confucius in fused silica. (a) The model and the (b) front, (c) left, (d) back, and (e) right sides of the sculpture. The details of the decorative pattern on the cloth, the right side of his face, and the left hand hanging behind his body are shown in the insets on the right-hand side of the images in (b), (c) and (d), respectively. Scale bar: 5 mm.

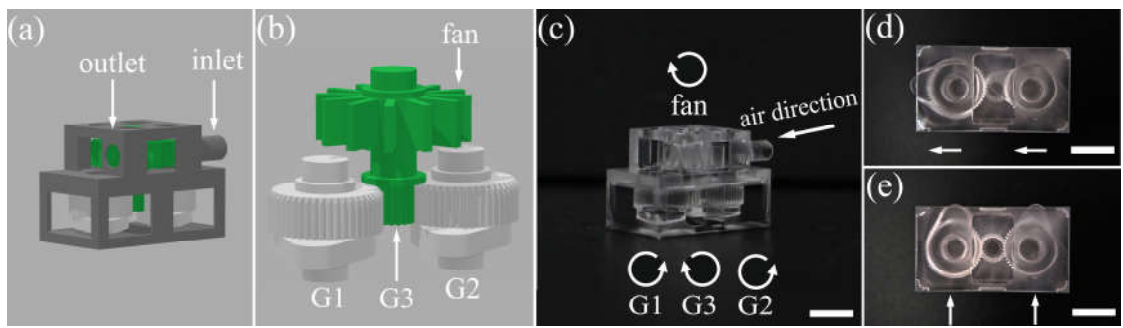

Fig. 4 A laser printed air turbine in fused silica. (a) The whole air turbine model. Inlet and outlet for air injection are indicated. (b) The interior of the turbine including turbine fan, a driving gear (G3) and two driven gears (G1 and G2). Each of G1 and G2 is connected with a cam. (c) Digital-camera captured image of the fabricated turbine. The air direction and the rotation direction of the fan, as well as the rotation directions of G1, G2 and G3 from a top view are all indicated by the curved arrows in (c). (d) The initial position of the two cams is pointing to the left as indicated by the two arrows. (e) Both the cams are rotated in a clockwise direction by 90° as a result of the injected air flow. Scale bar: 5 mm.

**Methods:**

**Experimental setup and fabrication process**: The experimental setup is schematically illustrated in Fig. S1. A Yb: KGW femtosecond laser (Pharos PH1-SP, Light Conversion) generated short pulses at 1030 nm wavelength with the highest single pulse energy at 1 mJ. The pulse duration can be continuously tuned from 0.19 to 10 ps with either a positive or a negative chirp. The repetition rate of the laser source can be adjusted between 1 kHz and 1 MHz. When performing the glass 3D micro-printing, the laser power was controlled using an attenuator. The laser beam diameter was first reduced to 2 mm in diameter using a telescopic system, and then passed through an acousto-optical modulator (AOM). The AOM was triggered by a radio frequency signal from the controller of the transition stage to offer a fast shutter speed less than 100 μs. Afterwards, the laser beam was expanded using a beam expander, and focused into the sample (i.e., a 55 mm-thick cube of fused silica) using a 5× objective lens (M Plan Apo NIR, Mitutoyo Corporation) with a numerical aperture (NA) of 0.14. The objective lens features a long working distance of 37.5 mm, enabling the fabrication of structures with heights up to 5 cm in fused silica. A 1D stage (ANT130-110-L-ZS, Aerotech Inc.) was used to translate the objective lens along Z direction to control the depth of the focus position in the glass. The glass sample was mounted on an XY motion stage (ABL15020WB and ABL15020, Aerotech Inc.) which controls the lateral motion of the sample with a scan speed up to 30 mm/s at a sub-500 nm positioning precision. Both the translation stages were controlled by the machine controller (A3200, Aerotech Inc.) and synthesized with the AOM. The combination of the high scan speed and the high positioning precision facilitates the rapid 3D micro-printing of large structures in glass as will be shown below.

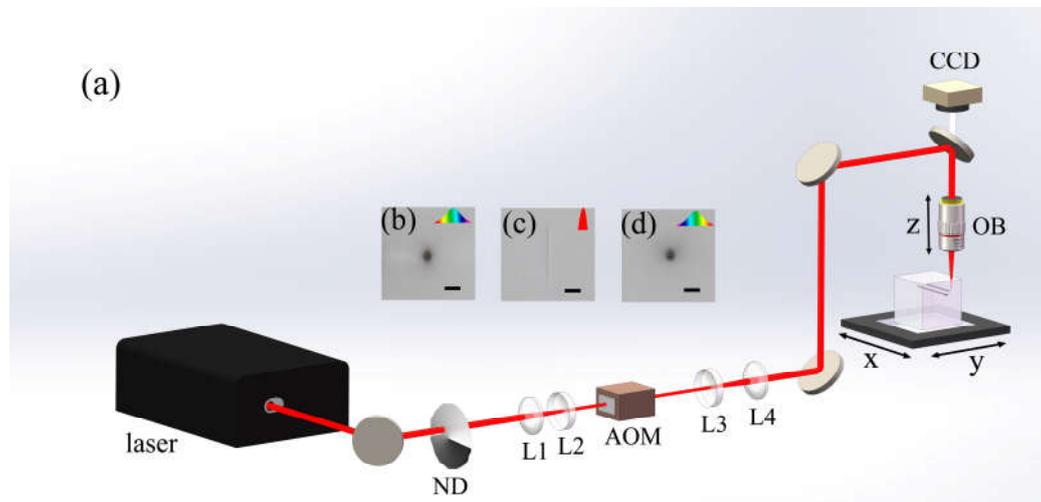

Figure S1 (a) Schematic of the experimental setup. ND, variable neutral density filter; L1 and L4, convex lens; L2 and L3, concave lens; AOM, acousto-optical modulators; CCD, charge coupled device; OB, objective lens. Cross-sectional view of optical micrographs of lines inscribed in fused silica with (b) positively chirped 10 ps laser pulses, (c) transform-limited 190 fs laser pulses and (d) negatively chirped 10 ps laser pulses. Scale bar, 25 μm.

**Data conversion:** The models used for 3D glass micro-printing were originally generated as stereolithography (STL) files. The STL were firstly covered by a cuboid frame, and then underwent a subtraction operation. Afterwards, the 3D models were sliced into horizontal planes with a fixed slice thickness. We scanned the laser focal spot in the sliced planes along the pre-designed paths layer by layer to produce the 3D structures. The scan process was performed from the bottom to the top of the glass. Typically, there are two scan strategies in the stereo lithography fabrication: the raster scan and the contour scan. In the former strategy, the whole volume of the structure should be scanned; whereas for the later one, the laser focal spot only scans along the contour profile of the 3D structure. In our experiment, since the irradiated part of the glass should be removed after wet etching to get the structure, the raster scan had been chosen.

**Chemical wet etching after the laser irradiation:** After the fused silica samples were selectively irradiated by the focused laser pulses as illustrated in the first two panels of Fig. S2, the samples were polished to remove the outer areas. The outer areas near the vertical sidewalls of the cubic sample cannot be sufficiently modified by the laser irradiation because of the distortion of the focal spot caused by the air-glass interface at the sidewalls. The polished samples were then immersed in a wet-etching bath of potassium hydroxide (KOH) with a concentration of 10 mol/L at a temperature of 90 °C for tens of hours as shown in the last two panels of Fig. S2. For example, it took in total ~72 hrs in the etching process to produce the 3.8 cm-tall Confucius sculpture, whilst the etching cycle time was reduced to ~36 hrs for producing the 1.8 cm-tall Einstein's head sculpture.

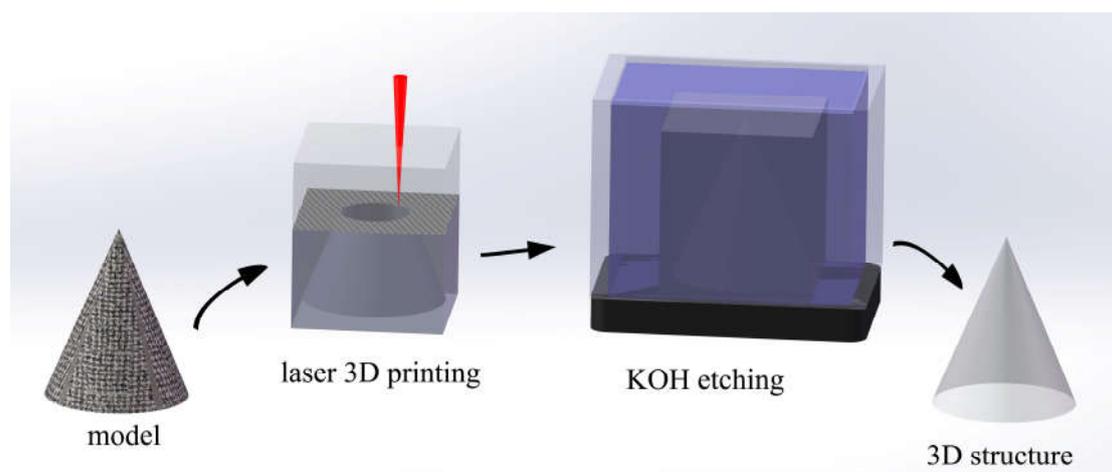

Fig. S2 The three major steps in the 3D glass micro-printing: Digitalization of the 3D model (first panel), scan of the laser beam along the predesigned paths to selectively modify the areas surrounding the 3D objects (second panel), removal of the irradiated materials with the chemical wet etching (third panel). The printed 3D structure is illustrated in the last panel.

**Optimization of the slice thickness and characterization of the surface quality**

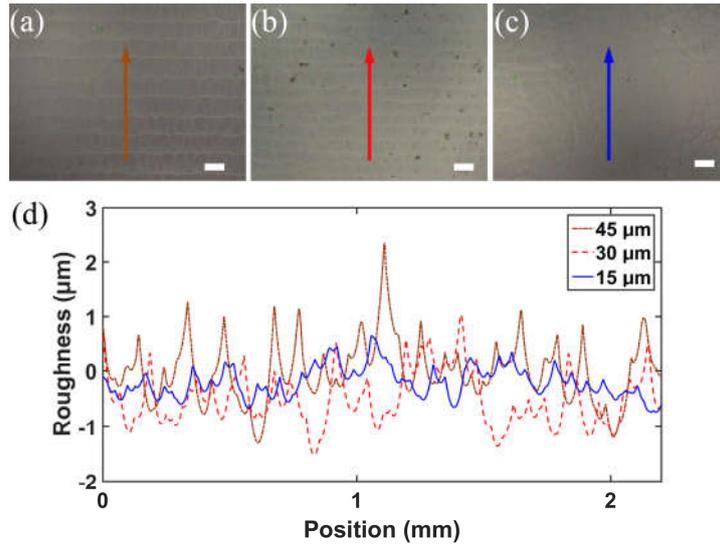

Fig. S3 Surface morphologies measured on samples written with the slice thicknesses set at (a) 45 μm; (b) 30 μm; and (c) 15 μm. (d) The measured 1D surface profiles of the samples in (a), (b) and (c) are shown by the green dotted, red dashed, and blue solid lines, respectively. Scale bar: 50 μm.

It is shown in Fig. 1 of the main text that an isotropic spatial resolution on the level of ~20 μm which is independent of the depth of the focal position can be obtained by loosely focusing the picosecond laser pulses into fused silica. However, for fabricating the 3D objects, chemical wet etching must be used, which leads to the degradation of the fabrication resolution. To determine the limit on the fabrication resolution along the longitudinal direction, we scanned the focused picosecond laser pulses in glass with different slice thicknesses and etched the samples in KOH to reveal the surface morphologies. As shown in Fig. S3(a-c), the surfaces produced with the slice thicknesses set at 45 μm and 30 μm show a clear laminar feature, owing to the fact that the thickness of each modification layer is less than both the slice thicknesses. When the slice thickness is reduced to 15 μm, the laminar feature disappears, leaving behind a uniform surface with a typical morphology given rise to by the FLICE. It is also shown in the 1D surface profiles of the samples measured with a surface profiler, as shown in Fig. S3(d), that periodic ripples oriented along the horizontal direction with a height between 1 μm and 2 μm and a height around 1 μm appear for the surfaces produced at the slice thicknesses of 45 μm and 30 μm, which agrees well with the observations in Fig. S3(a) and (b). However, when the slice thickness is reduced to 15 μm, the 1D profile along the cutting lines in Fig. S3(d) only shows random peaks without a significant periodicity.

In principle, the results in Fig. 3S indicates that a longitudinal resolution between 15 μm and 30 μm had been achieved with the picosecond laser 3D printing in glass, which agrees with the ~20 μm resolution as shown in Fig. 1(d) in the main text. Unfortunately, there is a physical limit so far in choosing such a small slice thickness in printing large 3D structures in thick fused silica. The problem is that when ultrafast laser pulses are focused into glass, the modified material will have a volume expansion which build up stress inside glass. For large-scale internal modifications, the stress will be high and can give rise to multiple cracks which spoil the printed 3D objects. To solve this issue, we need to reduce the filling ratio in performing the laser writing. This results in a lower fabrication resolution than that allowed by the voxel size of 3D laser writing. In the current experiment, the slice thickness is finally set at 50 μm to enable 3D printing of objects with heights up to 3.8 cm as shown in Fig. 3 in the main text.

## References

1. Gattass RR, Mazur E. Femtosecond laser micromachining in transparent materials. *Nat Photonics* **2**, 219-225 (2008).
2. Osellame R, Hoekstra HJWM, Cerullo G, Pollnau M. Femtosecond laser microstructuring: an enabling tool for optofluidic lab-on-chips. *Laser Photonics Rev* **5**, 442-463 (2011).
3. Sugioka K, Cheng Y. Femtosecond laser three-dimensional micro- and nanofabrication. *Appl Phys Rev* **1**, 041303 (2014).
4. Bellouard Y, *et al.* The Femtoprint Project. *J Laser Micro Nanoeng* **7**, 1-10 (2012).
5. Chen F, de Aldana JRV. Optical waveguides in crystalline dielectric materials produced by femtosecond- laser micromachining. *Laser Photonics Rev* **8**, 251-275 (2014).
6. Beresna M, Gecevicius M, Kazansky PG. Ultrafast laser direct writing and nanostructuring in transparent materials. *Adv Opt Photonics* **6**, 293-339 (2014).
7. Stuart BC, Feit MD, Rubenchik AM, Shore BW, Perry MD. Laser-Induced Damage in Dielectrics with Nanosecond to Subpicosecond Pulses. *Phys Rev Lett* **74**, 2248-2251 (1995).
8. Joglekar AP, Liu HH, Meyhofer E, Mourou G, Hunt AJ. Optics at critical intensity: Applications to nanomorphing. *P Natl Acad Sci USA* **101**, 5856-5861 (2004).
9. Liao Y, *et al.* Femtosecond laser nanostructuring in porous glass with sub-50 nm feature sizes. *Opt Lett* **38**, 187-189 (2013).
10. Couairon A, Mysyrowicz A. Femtosecond filamentation in transparent media. *Phys Rep* **441**, 47-189 (2007).
11. He F, *et al.* Fabrication of microfluidic channels with a circular cross section using spatiotemporally focused femtosecond laser pulses. *Opt Lett* **35**, 1106-1108 (2010).
12. Vitek DN, *et al.* Spatio-temporally focused femtosecond laser pulses for nonreciprocal writing in optically transparent materials. *Opt Express* **18**, 24673-24678 (2010).
13. Chu W, *et al.* Centimeter-Height 3D Printing with Femtosecond Laser Two-Photon Polymerization. *Adv Mater Technol-Us* **3**, 1700396 (2018).
14. Tan YX, *et al.* High-throughput multi-resolution three dimensional laser printing. *Phys Scripta* **94**, 015501 (2019).
15. Li X, *et al.* Polarization-insensitive space-selective etching in fused silica induced by picosecond laser irradiation. Preprint at https://arxiv.org/abs/1812.10661 (2018).
16. Shimotsuma Y, Kazansky PG, Qiu JR, Hirao K. Self-organized nanogratings in glass irradiated by ultrashort light pulses. *Phys Rev Lett* **91**, 247405 (2003).
17. Bhardwaj VR, *et al.* Optically produced arrays of planar nanostructures inside fused silica. *Phys Rev Lett* **96**, 057404 (2006).
18. Liao Y, *et al.* Direct laser writing of sub-50 nm nanofluidic channels buried in glass for three-dimensional micro-nanofluidic integration. *Lab Chip* **13**, 1626-1631 (2013).
19. Corbari C, Champion A, Gecevicius M, Beresna M, Bellouard Y, Kazansky PG. Femtosecond versus picosecond laser machining of nano-gratings and micro-channels in silica glass. *Opt Express* **21**, 3946-3958 (2013).

## Acknowledgements

The work is supported by Key Project of the Shanghai Science and Technology Committee (Nos. 18DZ1112700, 17JC1400400), National Natural Science Foundation of China (61590934, 11674340, 11734009, 11874375, 11822410, 11604351), National Key R&D Program of China (No. 2018YFB0504400) and the Strategic Priority Research Program of Chinese Academy of Sciences (No. XDB16000000).